\documentclass[10pt, conference, compsocconf]{IEEEtran}

\ifCLASSINFOpdf
\else
\fi

\usepackage{graphicx}
\usepackage[export]{adjustbox}
\usepackage{cite}
\usepackage{nameref}
\usepackage{url}
\usepackage{color,soul}
\usepackage{cprotect}

\begin{document}
%
\title{Interactivity and Transparency in Medical Risk Assessment with Supersparse Linear Integer Models}


\author{\IEEEauthorblockN{Hans-J\"{u}rgen Profitlich}
	\IEEEauthorblockN{Daniel Sonntag}
	\IEEEauthorblockA{German Research Center for Artificial Intelligence (DFKI)\\
		Technical Report\\
		66123 Saarbr\"{u}cken, Germany\\
		profitlich@dfki.de\\
		sonntag@dfki.de}
}

\newcommand{\charite}{Charit\'{e}}
\newcommand{\tbase}{TBase\textsuperscript\textregistered}

\maketitle

\begin{abstract}
Scoring systems are linear classification models that only require users to add or subtract a few small numbers in order to make a prediction. They are used for example by clinicians to assess the risk of medical conditions. This work focuses on our approach to implement an intuitive user interface to allow a clinician to generate such scoring systems interactively, based on the RiskSLIM machine learning library. We describe the technical architecture which allows a medical professional who is not specialised in developing and applying machine learning algorithms to create competitive transparent supersparse linear integer models in an interactive way. We demonstrate our prototype machine learning system   
in the nephrology domain, where doctors can interactively sub-select datasets to compute models, explore scoring tables that correspond to the learned models, and check the quality of the transparent solutions from a medical perspective.
\end{abstract}


\section{Introduction}

Risk scores are simple linear classification models where users assess risk by adding and subtracting a few  numbers. These methods are often used for criminological or medical applications because they allow users to make quick predictions without the use of statistics or a calculator. Such scoring systems are widespread and Wikipedia lists 42 "medical scoring systems", such as the Simplified Airway Risk Index for predicting difficult tracheal intubation. The score ranges from 0 to 12 points, where a higher number of points indicates a more difficult airway. A score of 4 or above indicate a difficult intubation.  Current medical scoring systems were mostly created manually by clinicians, where a panel of experts agrees on a model (see the CHADS2 score of Gage et al. \cite{Gage2001ValidationOC}, for example).

Despite the widespread use of medical scoring systems, there has been little to no work that has focused on machine learning methods to learn these models from data. The goal of the SLIM system \cite{DBLP:journals/ml/UstunR16} is to present a principled approach to learn risk scores by solving a discrete optimisation problem, namely the risk score problem.
Models should be fully optimised for feature selection, small integer coefficients, and operational constraints. The risk scores (in the medical domain) have to be rank-accurate, risk-calibrated, sparse, and use small integer coefficients. Of particular interest for interactivity is the fact that systems such as SLIM provide additional operational constraints to limit the model size, the range of coefficients or the maximal runtime to compute the model. 

With RiskSLIM (Risk-calibrated Supersparse Linear Integer Model), \cite{DBLP:conf/kdd/UstunR17} propose a new approach to build risk scores that are fully optimised for feature selection, small integer coefficients, and operational constraints without parameter tuning or post processing. They provide software to create optimised risk scores using Python and the CPLEX API (IBM ILOG CPLEX Optimizer is a tool for solving linear optimisation problems, commonly referred to as Linear Programming (LP) problems). Our work focusses on providing an interactive environment to test such improved models for applicability in the medical domain, to be used by doctors.

The RiskSLIM implementation forms the core of our Web-based interactive scoring system, and the new interactive environment should provide the following:
\begin{itemize}
	\item a user interface that should be used by a medical professional who is not specialised in developing and applying machine learning algorithms
	\item interactive user support to generate suitable data sets;
	\item visual metaphors and help organise data sets and corresponding models; and 
	\item evaluation support to check and compare the quality of different models.
\end{itemize}

Our work is aimed at building an architecture that overcomes the short-comings of batch machine learning scoring systems and enables the construction of scoring tables even for end users (cf. topics of interactive machine learning, see  iml.dfki.de). In our medical use case,  we rely on data from the \tbase\ data base of \charite\ Berlin \cite{schroeterTBase,lindemannTBase} which contains data about nephrology patients.
We define a workflow and implement wrapper modules that allows the user to create a project by defining a medical target and a list of input features, generate corresponding data sets, run the RiskSLIM algorithm, explore the resulting scoring table, and check the quality of models on different data sets. The whole process can be started and evaluated  via a single Web page interface. After explaining the project's background, we describe the system architecture  and the implementation of the Web-based user interface.

\section{Background}

We started with the clinical data intelligence project (KDI) \cite{Sonntag2016};  we transferred research and development results (R\&D) of the analysis of data which are generated in the clinical routine in a specific medical domain. We presented the project structure and goals, how patient care should be improved, and the joint efforts of data and knowledge engineering, information extraction (from textual and other unstructured data), statistical machine learning, decision support, and their integration into special use cases moving towards individualised medicine. In particular, we described some details of our medical use cases and cooperation with two major German university hospitals, one of them providing the nephrology data for medical risk assessment. 

Then we focussed on integrated textual information extraction and interactive facetted search applications in nephrology; these were KDI's first integration steps of complex and partly unstructured medical data into a clinical research database. Our main application was an integrated facetted search tool in nephrology based on automatic information extraction results from textual documents. 

Towards integrated decision support \cite{DBLP:conf/cbms/SonntagP17},  the two next logical steps were the visualisation of facetted search results and producing new results and insights with the help of machine learning.  \cite{DBLP:journals/artmed/SonntagP19} describes our steps to integrate complex and partly unstructured medical data into a clinical research database with subsequent decision support. Our main application is an integrated faceted search tool, accompanied by the visualisation of results of automatic information extraction from textual documents, and second,  case studies, illustrating how the application can be used by a clinician and which questions can be answered. For example, in nephrology we try to answer questions about the temporal characteristics of event sequences to gain significant insight from the data for cohort selection. However, the identification of correlations in medical data by faceted search has the potential to identify relevant groups of patients, diagnoses, parameters, and to identify correlations of influencing factors \cite{SCHMIDT201798}, but it cannot directly propose guidelines as decision support. Scoring systems however are linear classification models that allow us to infer not only influencing factors, but to build rule systems as machine-learned medical guidelines that are transparent and understandable by the medical experts. Examples are shown in the implementation section.  

Our domain is the nephrology department of the \charite\ Berlin.
The Web-based electronic patient record \tbase\ was implemented in a German kidney transplantation programme as a cooperation between the Nephrology of \charite\ Universit\"{a}tsmedizin Berlin and the AI Lab of the Institute of Computer Sciences of the Humboldt University of Berlin . Currently, \tbase\ automatically integrates essential laboratory data (9.9 million values), clinical pharmacology (237.000 prescribed medications), diagnostic findings from radiology, pathology and virology (146.000 findings), and administrative data from the SAP-system of the \charite\ (70.000 diagnoses, 25.000 hospitalizations). All these facts are potential input features for different models.

\section{System Architecture}

We implement a system architecture and user interface that offers all necessary functionalities in a pipeline:
\begin{enumerate}
	\item select a feature from some predefined set as the 'goal' or medical target;
	\item choose (from the remaining features) a set of features as input parameters;
	\item create data sets corresponding to these feature lists;
	\item call RiskSLIM to learn a model;
	\item aggregate an interactive scoring table representing this learned model; and 
	\item compute and display some evaluation and quality measures of the model like precision or recall. 
\end{enumerate}

\begin{figure*}
	\centering
	\includegraphics[width=0.9\textwidth]{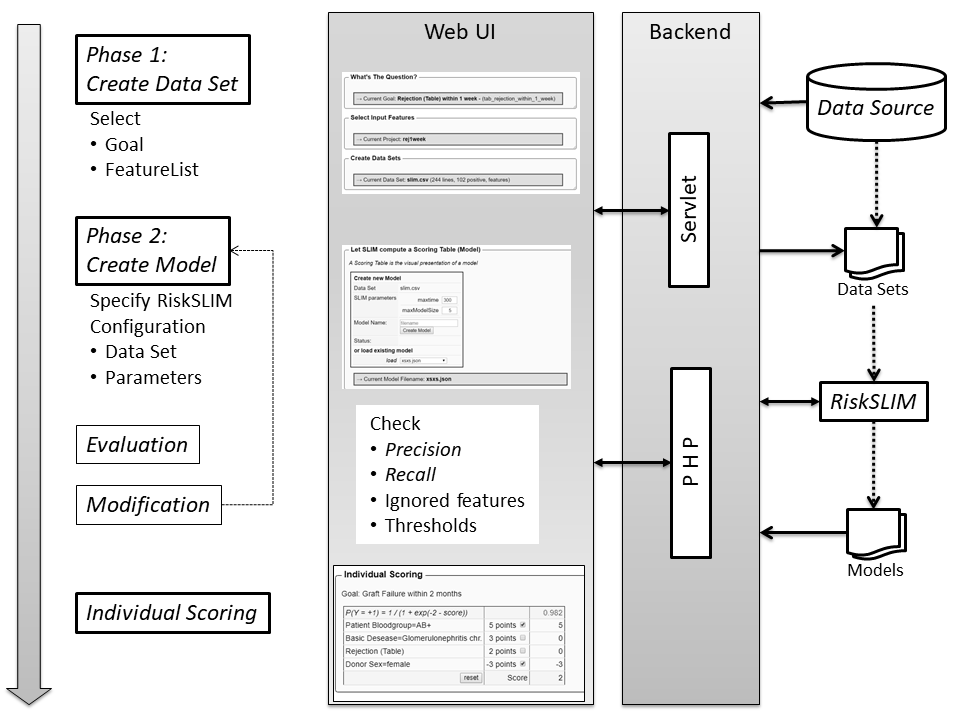}
	\caption{Workflow and Architecture}
	\label{fig:Architecture}
\end{figure*}

We define a \textit{project} to be the specification of a target plus a set of features (points 1 and 2 in the the list above). Data sets are always created relative to a project. This is necessary as the validation of a model can only be performed on data sets with the same structures as the data set the model was trained for. The workflow and architecture are shown in figure \ref{fig:Architecture}. The backend contains the user interface servlet and the PHP server accessing the data sources, the created input data sets, the RiskSLIM installation, and the learned models.

The RiskSLIM  machine learning library creates customised risk scores implemented in Python.
The implementation is available on GitHub\footnote{\url{https://github.com/ustunb/risk-slim}} and includes batch scripts that
\begin{itemize}
	\item read the input data from comma-separated files;
	\item set some configuration parameters;
	\item run the algorithm which outputs an array of integers representing the bias and coefficients for every input feature.
\end{itemize}

\begin{figure*}
	\centering
	\includegraphics[width=0.77\textwidth]{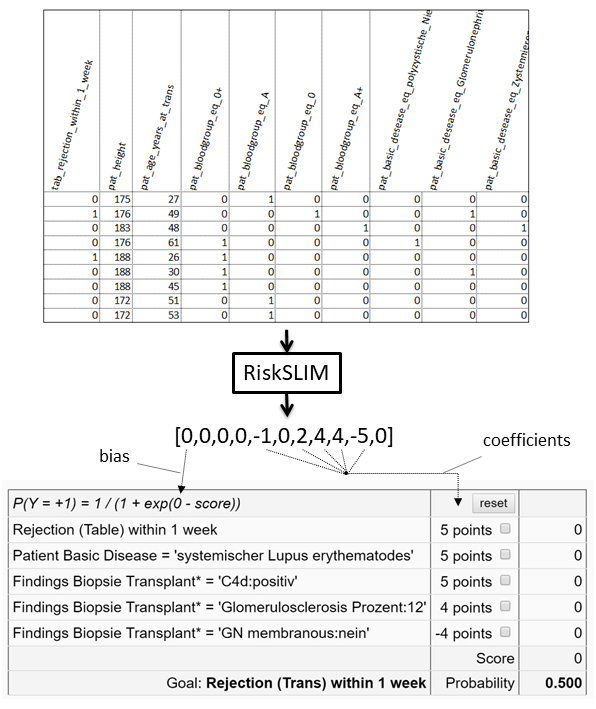}
	\caption{Input and output format of the RiskSLIM module and the visualisation of the computed model as a scoring table.}
	\label{fig:inputoutput}
\end{figure*}

The input files represent patient attributes with the target value in the first column, see the table at the top of figure \ref{fig:inputoutput} as an example: in the nephrology domain, an important target is the likelihood of a rejection of a kidney transplant within one week. For this purpose, the input features patient height, age at transplant, blood group, and basis diseases  are considered (after selection by the medical expert). 

The result of a RiskSLIM training cycle is  a vector of integers, representing the bias (first column/value) and the weight of each feature (remaining columns/values). Most of the input features have a weight of zero, the number of non-zero values (e.g., the size of the model) is one of the input parameters of the algorithm. In most cases we aim to create small models of sizes about five to seven features. 
The vector can be visualised as a scoring table (see bottom of figure\ref{fig:inputoutput}). The predicted risk for the defined target is computed by the formula
$$P(Y = +1) = 1 / (1 + exp(bias - score))$$
where $score$ is the sum of points related to the individual items of the scoring table.

In order to use the RiskSLIM software in clinical practice, we implement software modules to create data sets for training, testing and validating models and to interactively check the validity of models. 

\section{Implementation}

RiskSLIM is implemented as a Python package without any support for non-expert users. The Python package can only be run in a Python environment like PyCharm\footnote{\url{www.jetbrains.com/pycharm/}} or Anaconda Spyder\footnote{\url{anaconda.org/anaconda/spyder}} or from command line.
In the following we describe additional modules that embed this script into an environment consisting of a backend and a user front end that allow any clinician to generate scoring tables without any additional knowledge of the algorithms interfaces.

\begin{figure}
	\centering
	\includegraphics[width=0.7\columnwidth, frame]{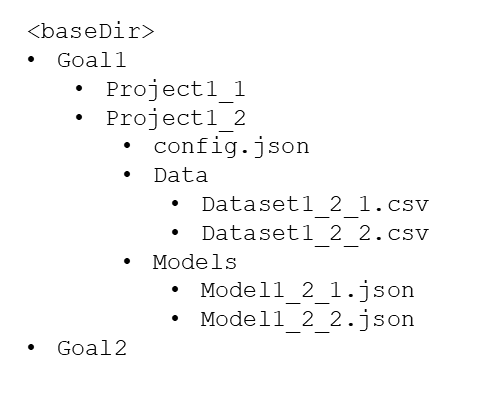}
	\caption{Folder structure}
	\label{fig:folderstructure}
\end{figure}

The necessary functionalities can be roughly divided into two groups:
1) communication with the data base and
2) operations on the file system.
We use the file system to organise projects and their corresponding data sets and models as illustrated in figure \ref{fig:folderstructure}.

\subsection{The Backend}

We implemented an additional Python script which can be given a data set and some runtime parameters. The script computes a model and saves the solution (and some additional data from the call) as a JSON file.

Data will be taken from the \tbase, a relational database, so we have to specify a pool of data the algorithm can use and how to access them.
As a first step we defined a list of features corresponding to patient meta data (like sex, age, weight, etc) and transplantation facts (like donor sex, donor age, previous transplants, diagnoses, events like rejections or graft failures within some time after the transplantation, etc).
For ever feature we specified 
\begin{itemize}
	\item an SQL statement 
	\item a readable label 
	\item a short text explaining what was used as a source for the feature value
	\item a flag telling whether all values are integers 
	\item a flag telling whether this is a multi-valued feature 
	\item a flag telling whether a feature can be chosen as a goal
\end{itemize}

Some of these features are straight-forward, simple attributes of patients, like age or blood group. But there are also more complex features like 'same sex of donor and patient' or the duration of dialysis before the transplantation was carried out (in years). There is also groups of features  representing the span of time between a transplantation and a graft failure (within one month, one year, two years, etc.) and similar relations. 
This initial list of features is more or less arbitrary and only serves as a starting point. As all system parts using features rely on this declarative representation additional features can easily be added without changing any other module.

To communicate with the \tbase\ we implemented a Java module and a servlet that offers the following functionalities:
\begin{itemize}
	\item getFeatureList: get a list of all defined features (with some of the additional facts mentioned above),
	\item createProject: create a new project according to a given configuration (a goal and a list of features),
	\item createDataSet: create a new data set given a project configuration\footnote{The data set can be restricted to a list of patients by supplying their ids.},
	\item validateModel: evaluate a model on a data set.
\end{itemize}

Data sets are matrices with a header row representing the feature names, a first column with the target value and trailing columns corresponding the remaining feature values (see figure \ref{fig:inputoutput}). When a data set is created from a list of features, the module has to ensure that all values are integers. We have to differentiate between three cases: 
\begin{enumerate}
	\item the feature has only one single value of type integer: the value is stored on a single column,
	\item the feature has $n$ non-integer values (e.g., blood group): the feature is represented by $n$ columns. The module creates a sparse vector of '0's and one single '1', column names are generated as '\textit{feature}EQ\textit{value1}', '\textit{feature}EQ\textit{value2}', etc.,
	\item the feature is multi-valued (e.g., biopsy results): analogously to case 2 a list of columns represents the different values, but this time more than one '1' is possible.
\end{enumerate}
The module automatically transforms the values read from the data base into the target format according to the flags specified for every feature.

A PHP script is used for operations on the file system (the projects, data sets and models are organised using an appropriate folder structure, see figure \ref{fig:folderstructure}) and to call the main RiskSLIM script:
\begin{itemize}
	\item getProjects: get a list of projects already defined for a (target) feature,
	\item loadProject: get the configuration of a project (mainly the list of features),
	\item getDataSets: get a list of generated data sets for a project,
	\item getModels: get a list of all models computed for a project,
	\item loadModel: load the data of a specific model,
	\item createModel: call RiskSLIM to compute a model.
\end{itemize}

\subsection{The Front End}

The complete functionalities necessary to control the processes are bundled in a single user interface in one Web page. 
The Web user interface was built using AngularJS 1.3\footnote{www.angularjs.org}, a JavaScript-based open-source Web application framework mainly maintained by Google to address challenges encountered in developing single-page applications. It aims to simplify the development of such applications by providing a framework for client-side model-view-controller (MVC) and model-view-view-model (MVVM) architectures.

The Web page supports the user in the execution of all working tasks and steps as shown in figure \ref{fig:Architecture}. In every step (define a goal, a feature list, a data set, create a model) the user can create a new item or choose from existing and compatible items.

The first phase consists of defining a project, that is, to specify a goal (a single feature) and a list of input features, which could be relevant for the goal. These specifications are then used to generate matching data sets. 

\begin{figure*}
	\centering
	\includegraphics[width=0.8\textwidth]{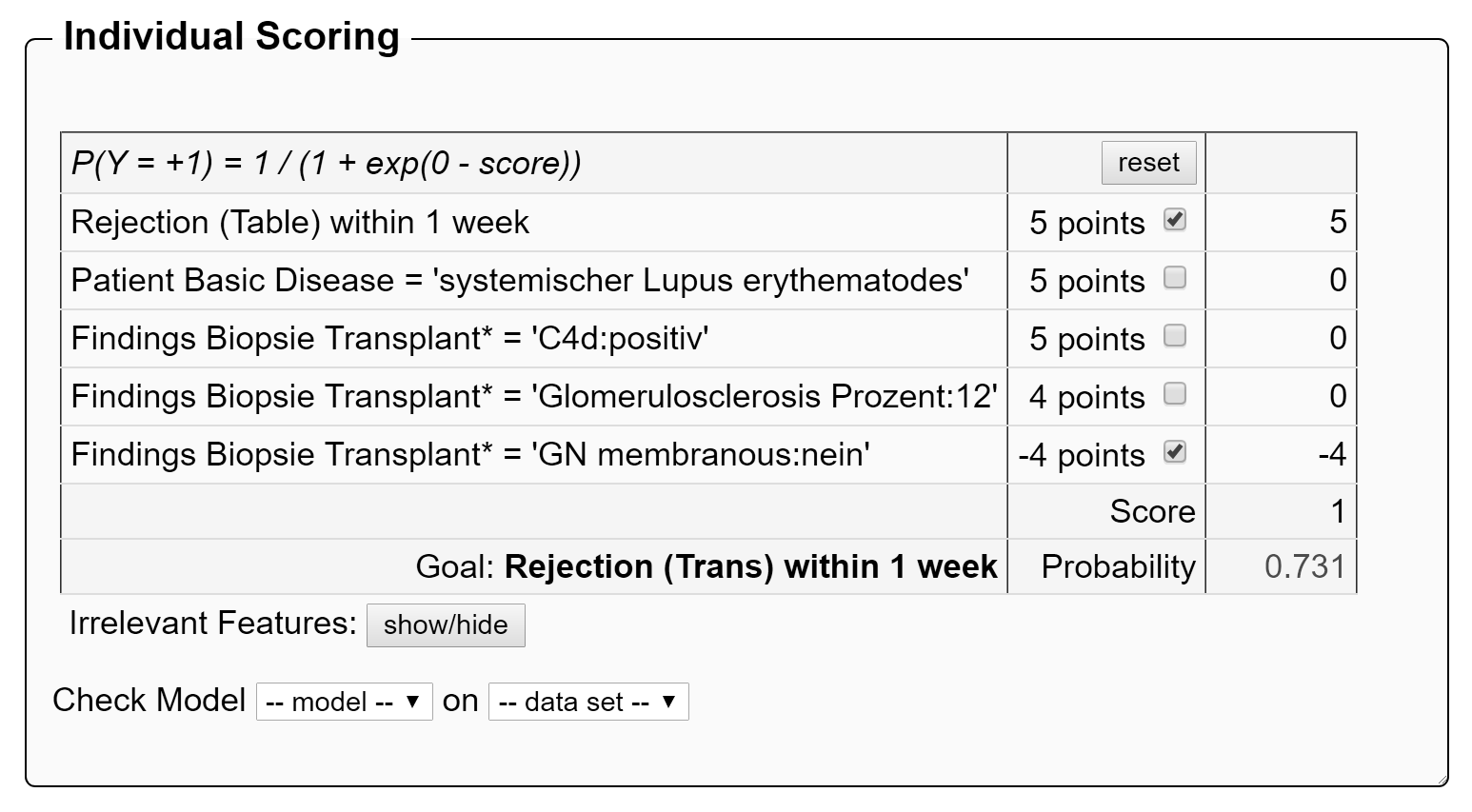}
	\caption{Interactive scoring table}
	\label{fig:scoringtable}
\end{figure*}

In the next phase, a call to the RiskSLIM algorithm can be initiated after specifying some simple parameters (runtime, model size). The resulting model is visualised as an interactive scoring table representing the computed most important features with their coefficients and showing the resulting risk scores for the defined target feature, see figure \ref{fig:scoringtable}.

\begin{figure}
	\centering
	\includegraphics[width=1.0\columnwidth]{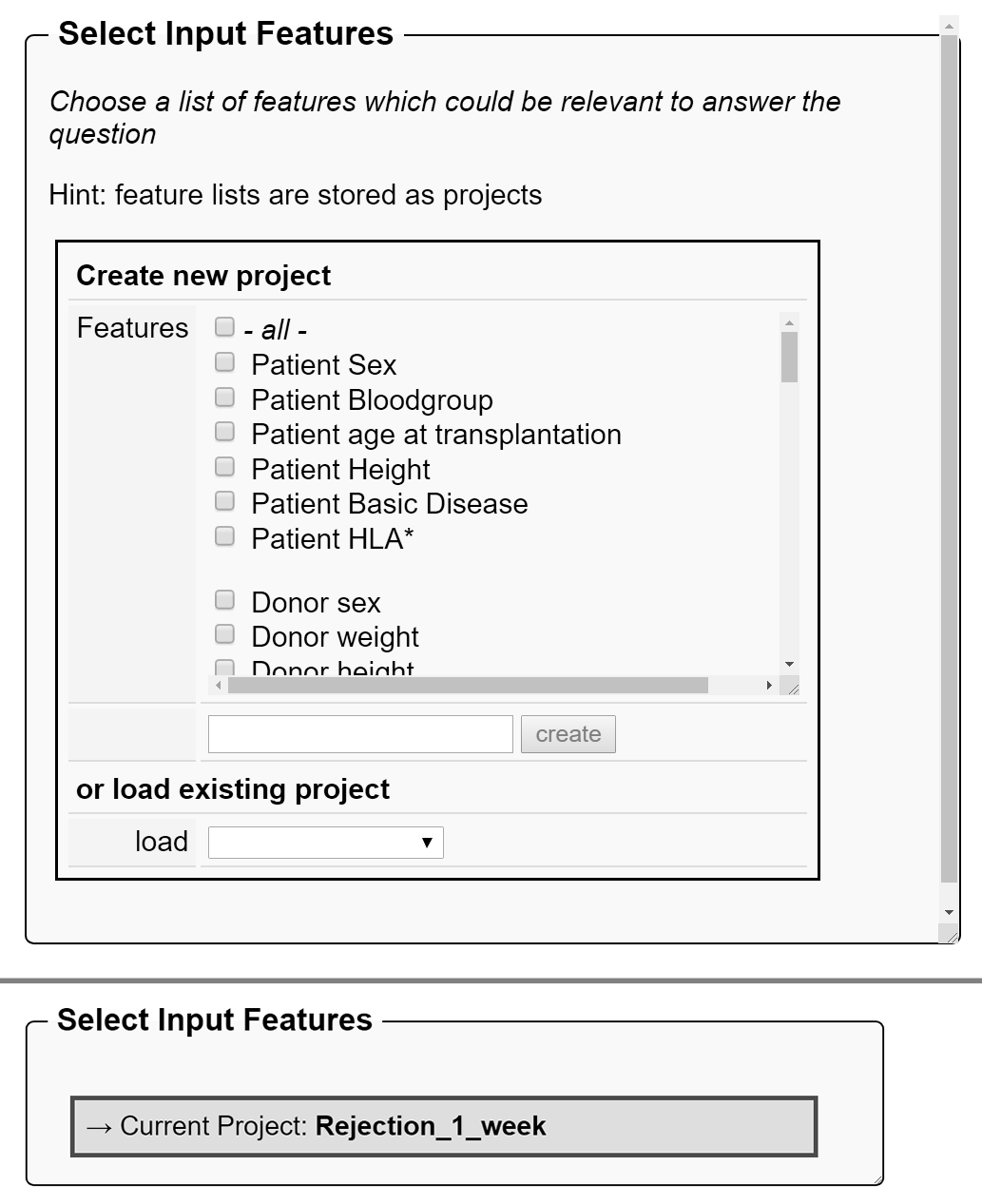}
	\caption{Top: Block 'Input Features': specifying the complete list of features; bottom: closed block showing current value (name of the project)}
	\label{fig:choosefeatures}
\end{figure}

As this  process can easily get confusing for a non-technical expert, we represent each step as a block with a heading line, some explanations, options to choose from and a value representing the result of this step. A block can be opened or closed, showing only the header and the current value when closed. Figure \ref{fig:choosefeatures} shows the open block for specifying the list of features for a project (top) and the same block closed showing only the current value (bottom).

If the user wants to change the options in a block, he or she just has to click the block showing the result to re-open the block. By this we can support the user to focus on the current step in the workflow by providing just the information needed at this moment and, at the same time, offering all flexibility needed. This progressive disclosure model is an interaction design technique often used in human computer interaction. It  helps to maintain the focus of a user's attention by reducing clutter, confusion, and cognitive workload \cite{Nielsen:2006:PWU:1137780}.

\begin{figure*}
	\centering
	\includegraphics[width=0.8\textwidth]{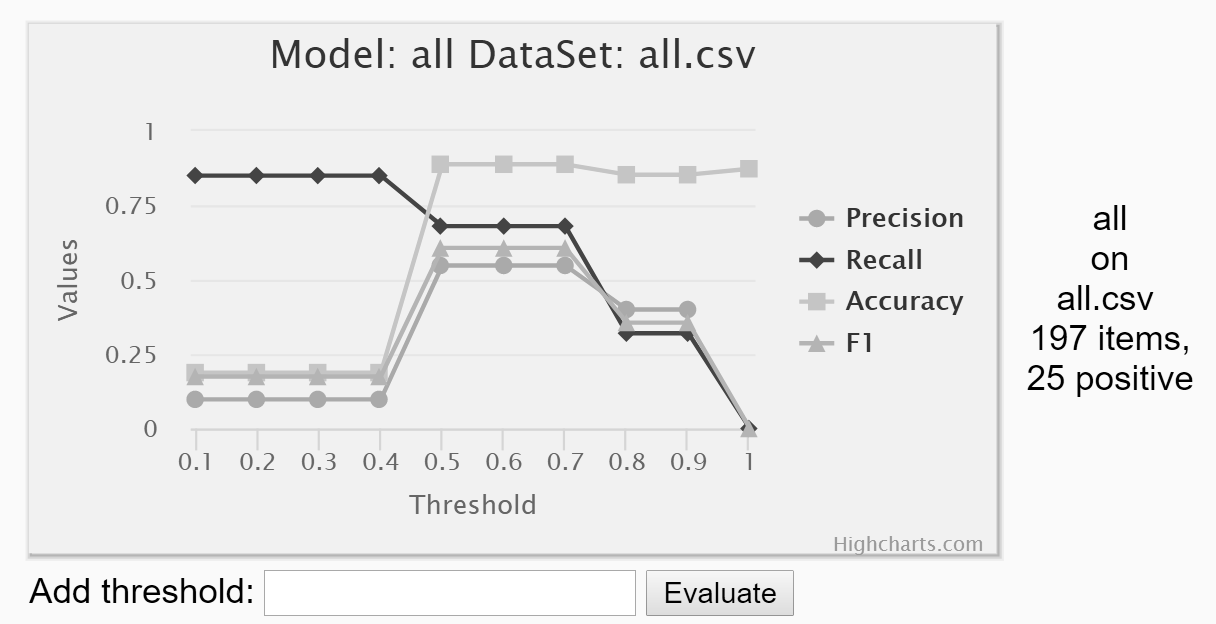}
	\caption{Validation diagram}
	\label{fig:diagram}
\end{figure*}

\begin{figure*}
	\centering
	\includegraphics[width=0.8\textwidth, frame]{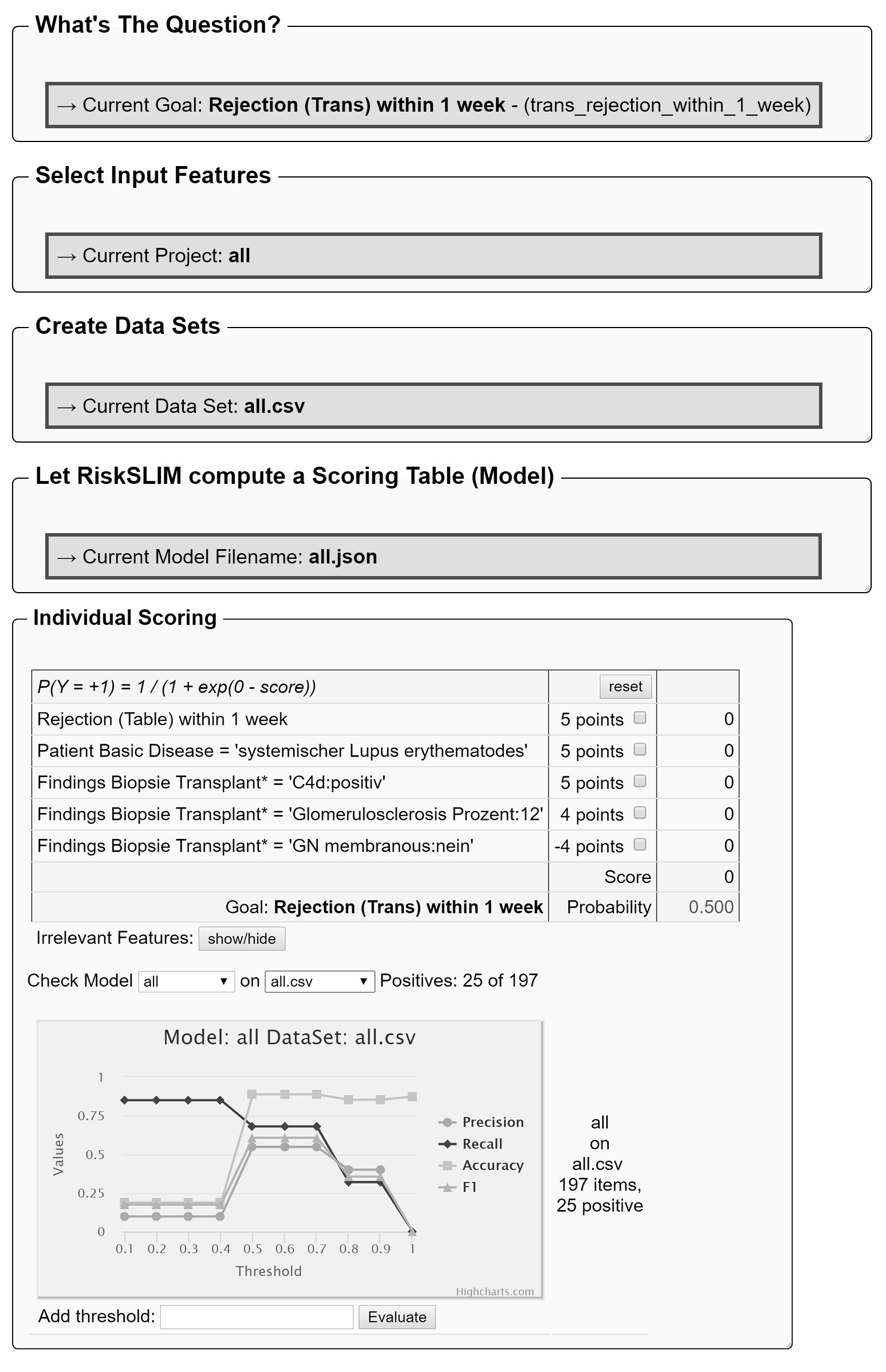}
	\caption{Web page showing the complete workflow}
	\label{fig:complete}
\end{figure*}

As a last step we added some functionalities to validate the quality of solutions.
As the model serves to compute the probability of a target feature, its value depends on a chosen threshold above which the target is assumed to be true. Two menus allow us to select a model and a data set (belonging to this project) for validation. After two values are chosen, a graph shows some quality measures (i.e., precision, recall, accuracy, and F1) for different possible thresholds (see figure \ref{fig:diagram}). Additional thresholds can be entered and are automatically added to the diagram. When other models or data sets are chosen, their diagram is appended (the previous diagrams remain visible) to allow for a direct comparison of different solutions.
Figure \ref{fig:complete} shows the complete Web page at the end of the workflow.

\section{Conclusion and Outlook}

We presented an intuitive and user friendly work environment for Medical Risk Assessment with Supersparse Linear Integer Models. It enables medical doctors to generate their own scoring tables based on the RiskSLIM library. 

The complete workflow from selecting a goal to the generation of data sets, the computation of models, the interactive testing of scoring tables up to the validation of different solutions can be performed from one Web page without any additional knowledge. 
Currently the software is being deployed and tested at \charite\ Berlin by clinicians to check its utility for \tbase\ and its usability. The evaluation of the resulting scoring tables  can only be performed by medical doctors with the knowledge about a reasonable selection of features to compute the probability of target features, and the interpretation of the scoring systems themselves, which we try to interpret as clinical guidelines. 

Additional steps can be included in the workflow in the future, e.g., data engineering tasks like handling outliers or missing values, the binning of data, automatic partitioning of data sets for training, testing and validation, or automatic cross-validation, or including medical ontologies \cite{article09}. In addition, the set of features can be increased by including more potentially relevant attributes of patients or transplantations by using concepts of interactive machine learning (iml.dfki.de) and intelligent user interfaces \cite{DBLP:journals/corr/Sonntag17} in multimodal environments for the doctor \cite{sonntag2009supporting,Oviatt:2017:HMI:3015783,Sonntag:2019:MHS:3233795.3233808}.

\section*{Acknowledgements}
This research is part of the project ''clinical data intelligence'' (KDI) which is founded by the Federal Ministry for Economic Affairs and Energy (BMWi), and EIT Digital Skincare founded by Horizon 2020. Out thanks go out to Klemens Budde and Danilo Schmidt for providing access to  \tbase. 

\bibliographystyle{apalike} 
\bibliography{slim2}

\end{document}